\documentclass[floatfix,aps,physrev,twocolumn,unsortedaddress,showkeys,10pt]{revtex4-2}

\usepackage{graphicx}
\usepackage{siunitx}
\usepackage{physics}
\usepackage{bbold}
\usepackage[normalem]{ulem}

\usepackage{verbatim}

\begin{document}

%TC:ignore
% \detailtexcount{main}
\title{\textbf{Heralded one-, two- and three-photon states from \\ waveguided parametric down-conversion}}

\author{Daniel Borrero Landazabal}
\email{daniel.borrerolandazabal@dlr.de}
\author{Kaisa Laiho}
\affiliation{German Aerospace Center (DLR e.V.), Institute of Quantum Technologies, Wilhelm-Runge-Str. 10, 89081 Ulm, Germany}

\date{\today}

\begin{abstract}
The manifold coincidences and singles counting provides a resource-saving quantum-optics analysis tool, since it is appropriate even in the presence of heavy experimental imperfections. Here, we prepare cross-polarized twin beams in the telecommunication wavelength range via parametric down-conversion in a periodically-poled KTiOPO$_4$ waveguide and herald photon-number states up to three photons. First, we show that the single-click probability is a versatile tool not only for extracting the state's mean photon number but also for sampling values of the moment generating function being the core behind any quantum optical state. Second, we measure values of the normalized factorial moments of photon number, $g^{(m)}_{\text{h}}$, up to the order $m = n+1$ for the heralded $n$-photon state. These normalized photon correlations provide an expedient method for examining the higher-order non-classicality of light by violating the condition $g^{(m+1)}_{\text{h}} \ge g^{(m)}_{\text{h}} \ge 1$.

\end{abstract}
%TC:endignore

%\keywords{}

\maketitle

\emph{Introduction.} 
In the past, different approaches have been taken to produce and detect the peculiarities of non-classical light \cite{migdall2013single,olivares2021introduction}. Probably one of the most fascinating class of quantum light is formed by the  number states denoted as $\ket{n}$ ($n \in \mathbb{N}$) \cite{achilles2006direct, waks2006generation}. The process of parametric down-conversion (PDC), which generates photons in pairs that are called signal ($s$) and idler ($i$) \cite{burnham1970observation,rarity1992quantum}, suits well for the heralded generation of non-classical states of light, like the multi-photon number states \cite{tiedau2019scalability, zapletal2021experimental}. Due to the high photon-pair generation rate and the capability to herald close-to-pure quantum states of light, the periodically-poled KTiOPO$_4$ (PP-KTP) has become a well-established platform for this purpose \cite{eckstein2011highly,Bruno2014pulsed, Chen2017efficient,faleo2026optimized,barcons2026pure}. 

The measurement of the second-order Glauber correlation $g^{(2)}_{\text{h}}$ lies at the core of photon counting measurements \cite{meyer2020single}. Even though $g^{(2)}_{\text{h}} <1 $ is a valid proof of non-classicality \cite{harder2016single}, it is rather hard to discern the behavior of the multi-photon contributions involving photon numbers $\ge2$  from this measurement. Therefore, one must search for alternatives for retrieving their behavior \cite{hotter2026quantum}. For example, non-classicality criteria have been developed for directly addressing the PDC multi-photon contributions \cite{waks2004direct, waks2006highly}.
Alternatively, such criteria can be formulated via the joint normalized photon correlations between signal and idler, which offer simpler and more direct verification methods  \cite{avenhaus2010accessing, bohmann2019detection}.

Regarding the generation of free-traveling optical number states, numerous methods exist for retrieving hints of their non-classicality \cite{lvovsky2001quantum, ourjoumtsev2006quantum, laiho2010probing, nehra2019state, sonoyama2024generation, fenwick2026ultrafast}.
Furthermore, complex non-classicality criteria have been formulated for them based on matrices of the normalized photon correlations \cite{sperling2017identification}.
Instead, we make use of the normalized higher-order correlation functions of the heralded states to corroborate the violation of $g^{(m+1)} \ge g^{(m)} \ge 1$ that applies to classical light \cite{klauder2006fundamentals}. Most importantly this criteria provides an intuitive way for verifying the degree of non-classicality. These moments can handily be evaluated from raw clicks of coincidences and singles. They typically are loss- and realization-independent \cite{laiho2022measuring}, which makes them easy to implement for any quantum light source. Moreover, we extend our treatment to the moment generating function, which is a rich  element for the state tomography \cite{barnett2002methods}. The overwhelming similarity between the moment generating function and the expectation values of click-detection allows one to sample it in a simple manner.

In this work, we prepare optical number states via PDC emission from a PP-KTP waveguide in the telecommunication wavelength range by heralding $k$ detection events $(k = 1,2,3)$ in idler. In the following, we denote the heralded state in signal with $\ket{k}$. We start with the conventional characterization of the photon pairs by examining the coincidences-to-accidentals ratio (CAR) and the Klyshko efficiencies. Thereafter, we measure the click-probabilities for extracting the mean photon numbers of the heralded states and to sample values of their moment generating function. Via manifold coincidences and singles counting, we further  directly extract the values of $g_{\text{h}}^{(m)}$ with $m \leq k+1$ for the heralded states and investigate the violation of the higher-order non-classicality criteria. Altogether, we provide a resource saving approach for a precise multi-photon state classification in the presence of experimental imperfections.

\begin{figure}
    \centering
    \includegraphics[width=0.7\linewidth]{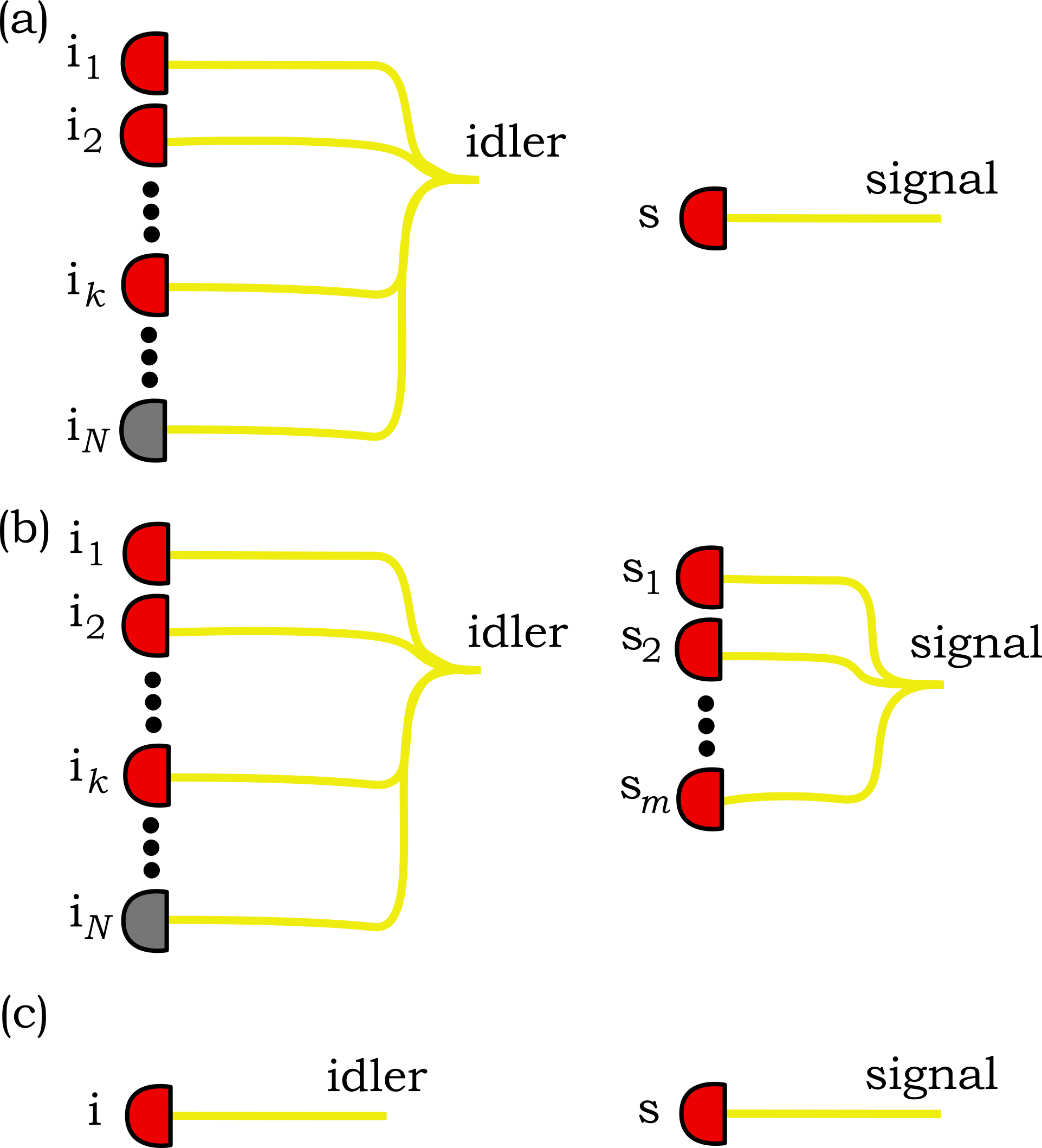}
    \caption{Sketches of the array-equivalent forms of the used detection with maximally $i_y$ $(y = 1,\dots N)$ bins in idler and $s_x$ $(x = 1,\dots m)$ bins in signal. The red detectors represent a detection event, while the gray ones are exposed by optical vacuum. For more details see the main text.}
    \label{fig:model}
\end{figure}

\emph{Theoretical treatment.~} We start with a theoretical treatment of our experimental arrangement.
In order to experimentally recover the mean photon number of the heralded states, we consider the probability of measuring a single click in signal by heralding $k$ simultaneous detection events in idler, as depicted in Fig.~\ref{fig:model}(a). Thereafter, to measure the normalized higher-order factorial moments of  the heralded states, we also evaluate the probability of counting $m$-fold coincidences in signal, while heralding $k$ detection events in idler, as shown in Fig.~\ref{fig:model}(b). Additionally, to measure the Klyshko's efficiencies and the CAR we employ the arrangement presented in Fig.~\ref{fig:model}(c). 

For extracting the mean photon number of the heralded states we calculate the single click-detection probability in the signal arm via the projection to \cite{landazabal2026validating}
\begin{equation}
    \hat{O}^{\text{s}} = \mathcal{I} -  \sum^{\infty}_{n=0}\left(1-\mu_{\text{sc}}\right)^{n} \ket{n}_{\text{s~s}}\bra{n} \, ,
\label{eq:POVM_signal}
\end{equation}
with $\mu_{\text{sc}}$ being the detection efficiency and $\mathcal{I}$ the identity.
Thus, the probability of detecting a single-click in signal is given by
\begin{equation}
    \Tr_\text{s}\{\hat{\rho}^{(k)}_{\text{s}}\hat{O}^{\text{s}}\} =  \mu_{\text{sc}} \expval{\hat{n}}^{(k)}_\text{s} + \mathbb{O}(2) \, ,
\label{eq:Tr_s}
\end{equation}
where $\hat{\rho}^{(k)}_{\text{s}}$ is the density matrix of the heralded state and $\mathbb{O}(2)$ accounts for a correction with terms that are proportional to at least $\mu_{\text{sc}}^2$ and thus negligible in our experiment. Experimentally, the probability in Eq.~(\ref{eq:Tr_s}) can be evaluated through
\begin{equation}
     \Tr_\text{s}\{\hat{\rho}^{(k)}_{\text{s}}\hat{O}^{\text{s}}\} = \frac{C(\text{i}_1,...,\text{i}_k, \text{s})}{S_{\text{i}_k}} \equiv \mathcal{P}_k\, ,
\label{eq:Tr_s2}
\end{equation}
in which  $S_{\text{i}_k}$ is the total amount of heralding events with $k$ simultaneous detection events in idler and $C(\text{i}_1,...,\text{i}_k, \text{s})$ is the total amount of coincidence events between one detection event in signal and $k$ simultaneous detection events in idler (cf.~Fig.~\ref{fig:model}(a)).
By combining Eqs.~(\ref{eq:Tr_s}) and (\ref{eq:Tr_s2}), the mean photon number of the $\ket{k}$-state is extracted by
\begin{equation}
  \expval{\hat{n}}_\text{s}^{(k)} = \frac{\mathcal{P}_k}{\mu_{\text{sc}}} \, .
 \label{eq:mean_exp}
\end{equation}
Furthermore, we note the remarkable similarity between Eq.~(\ref{eq:POVM_signal}) and the operator related to the moment generating function \cite{barnett2002methods}
\begin{equation}
    \hat{M}(\eta) = \sum_{n=0}^{\infty}(1-\eta)^n\ket{n}\bra{n},
\label{eq:Moment_Function}
\end{equation}
in which $0\le\eta\le 2$. Indeed, the value of the moment generating function at the point $\eta = \mu_{\text{sc}}$ can be directly sampled by evaluating the expectation value $\big < \hat{M}(\mu_{\text{sc}})\big>=\big <\mathcal{I}\big>-\big<\hat{O}^{\text{s}}\big> = 1 -\mathcal{P}_k.$ 

The normalized $m$-th order factorial moment for the $\ket{k}$-state, $g^{(m)}_{\text{h}} (\ket{k})$, can be retrieved from the m-fold coincidences in signal after heralding $k$ simultaneous detection events in idler (cf.~Fig.~\ref{fig:model}(b)). 
At low detection efficiencies, these moments can be evaluated as \cite{avenhaus2010accessing}
\begin{equation}
    g^{(m)}_{\text{h}} \left(\ket{k}\right) = \frac{\left(S_{\text{i}_k}\right)^{m-1} C(\text{i}_1,...,\text{i}_k, \text{s}_1,...,\text{s}_m)}{C(\text{i}_k,\text{s}_1) C(\text{i}_k,\text{s}_2)\cdots C(\text{i}_k,\text{s}_m)} \, ,
\label{eq:gmh_exp}
\end{equation}
where $S_{\text{i}_k}$ is the total amount of heralding events with $k$ simultaneous detection events in idler, $C(\text{i}_1,...,\text{i}_k, \text{s}_1,...,\text{s}_m)$ is the amount of coincidences between $k$ simultaneous detection events in idler and $m$ simultaneous detection events in signal, and $ C(\text{i}_k,\text{s}_{x})$ ($x = 1,\dots m$) is the amount of coincidences between $k$ simultaneous detection events in idler and a single detection event in the bin $\text{s}_{x}$ in signal.

For completeness, the  Klyshko's efficiencies are retrieved from coincidence counting via
\begin{equation}
    \mu_{\text{sc/ic}} = \frac{C(\text{i},\text{s})-C_{\text{acc}}}{S_{\text{i/s}}} \, ,
\label{eq:efficiencies}
\end{equation}
where $C(\text{i},\text{s})$ are the coincidences between idler and signal and $S_{\text{i/s}}$ the single's detection events in idler or signal (c.f Fig.~\ref{fig:model}(c)). Further, we approximate the accidentals as $C_{\text{acc}}=S_{\text{i}} S_{\text{s}} / R_{\text{pump}}$ with $R_{\text{pump}}$ being the pump repetition rate. Similarly, we extract the value of CAR experimentally via $\text{CAR} = C(\text{i,s})/C_{\text{acc}}$.

\begin{figure}[t]
    \centering
    \includegraphics[width=\linewidth]{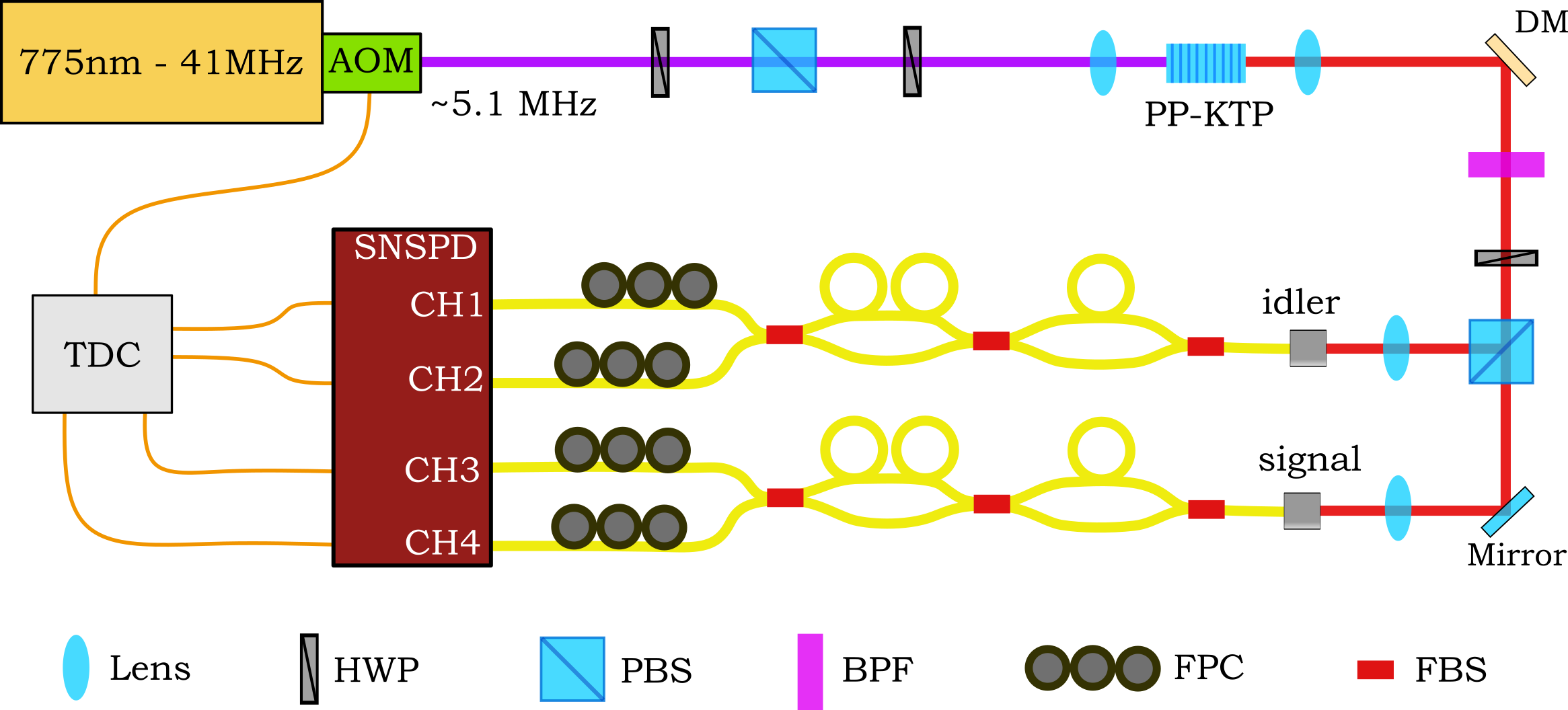}
    \caption{Experimental arrangement for the generation and characterization of heralded states. For Abbreviations see the main text.}
    \label{fig:setup}
\end{figure} 
\emph{Experimental arrangement.~} Our experimental arrangement is shown in Fig.~\ref{fig:setup}.  The pump light at $\SI{775}{\nm}$ is created by a pulsed laser system with \SI{41}{MHz} repetition rate and $\SI{350}{\fs}$ pulse duration. The pump light is sent into a fiber-integrated acousto-optic modulator (AOM), which reduces the repetition rate by a factor of eight leading to $R_{\text{pump}}\approx \SI{5.1}{\mega\hertz}$ and a repetition time of $\SI{195}{\ns}$. The pump power is controlled with a half-wave plate (HWP) placed in front of a polarizing beam splitter (PBS). The pump light propagates through another HWP before it is coupled to the PP-KTP waveguide. An aspheric lens with $\SI{6.24}{\mm}$ focal length is used for focusing it into the waveguide having a length of $\SI{11}{\mm}$ and $\numproduct{4x4}~\unit{\micro\meter^2}$ cross-sectional area. Afterwards, the type-II PDC emission, that is, the created signal and idler photons are cross-polarized, is collimated with an aspheric lens having a $\SI{3.1}{\mm}$ focal length. The pump beam is separated from signal and idler with a dichroic mirror (DM). Another  $\SI{12}{\nm}$-bandpass filter (BPF) centered at $\SI{1550}{\nm}$ is employed for spectral filtering. Signal and idler are sent through a HWP and separated in a PBS. Each of them is coupled into single-mode fibers by using aspheric lenses with $\SI{5.5}{\mm}$ focal lengths. 

\begin{figure}[t]
    \centering
    \includegraphics[width=\linewidth]{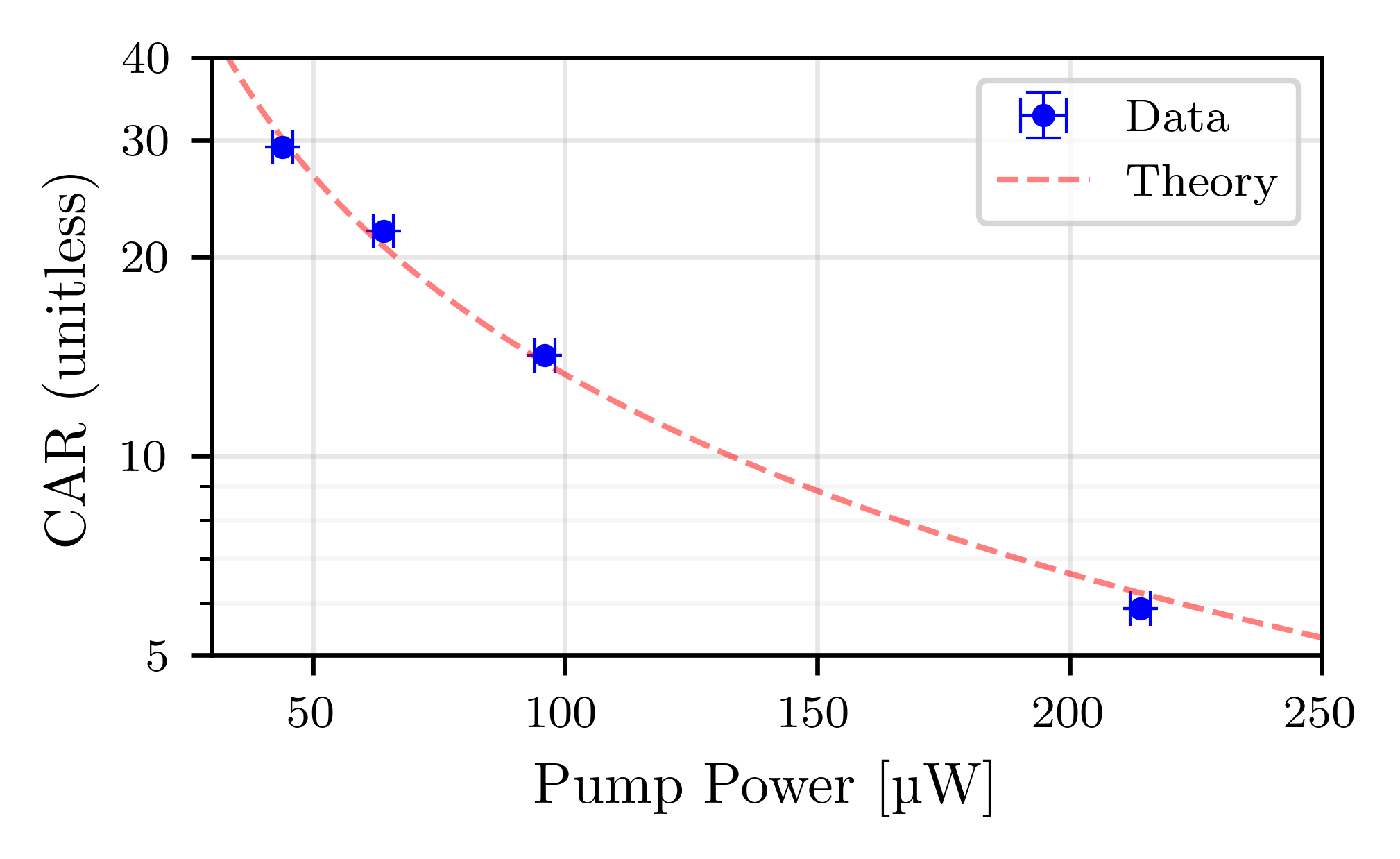}
    \caption{The measured CAR values in terms of the PDC pump power. The red dashed line presents a theoretical fit.}
    \label{fig:Car_power}
\end{figure}

\begin{table}[b]
\centering
\begin{tabular}{ccccc}
CAR & \SI{5.9}{} & \SI{14.2}{} & \SI{21.9}{} & \SI{29.3}{} \\
\hline \vspace{-3mm} \\ \hline \vspace{-2mm}\\
$ \mu_{\text{sc}} (\%)$ & \SI{16.2}{} & \SI{16.8}{} & \SI{16.3}{} & \SI{16.7}{} \\
\hline \vspace{-3mm} \\ \hline \vspace{-2mm}\\
$ \mu_{\text{ic}}(\%)$ & \SI{21.4}{} & \SI{21.9}{} & \SI{24.4}{} & \SI{24.1}{} 
\end{tabular}
\caption{Measured Klyshko's efficiencies for signal and idler at each value of the CAR approximated to the first decimal place. The statistical errors in these values are smaller due to the large ensembles of collected events.}
\label{Tab:efficiency}
\end{table}

For the manifold coincidence counting, we implement time-multiplexed detection (TMD). Our TMDs consist of three $50/50$ fiber-optic beam splitters (FBS) adjunct together with fibers of different lengths and connected via fiber polarization controllers (FPCs) to superconducting nanowire single-photon detectors (SNSPDs). Thus, an incoming pulse in the signal or idler beam paths is separated to a train of four pulses in any SNSPD channel (CH) with a relative delay of around $\SI{50}{\ns}$ between those temporal bins. In total, there are eight temporal bins in each signal and idler.
The electric output of the SNSPD channels are connected to a time-to-digital converter (TDC)  for counting the statistics. The count rates are measured with respect to a trigger delivered by the AOM. We further employ a detection window of $\SI{0.5}{\nano\second}$ in each individual temporal bin in order to suppress the effect of spurious counts. The measurements for heralding $k$-clicks $(k = 1,2,3)$ are performed simultaneously. In order to collect enough statistics, the measurement were run twice for around 12 hours at each pumping power. The pumping powers were ranged from \SI{44}{\micro W} to \SI{214}{\micro W}. At the lowest pump power the ensembles of collected statistics include $1\times 10^{9}$, $5\times 10^{6}$, and $7\times 10^{3}$ heralding events for $k=1,2,3$, respectively. 

\emph{Results and discussion.~} First we extract the conventional characteristics of the underlying photon-pair generation process. For this purpose, in Fig.~\ref{fig:Car_power} we  present the values of the CAR in terms of pump power. The CAR is an appealing calibration parameter, which follows an inverse proportionality to the pump power. The measured CAR values vary approximately from $5.9$ to $29.3$ such that we remain in the region of strong photon-pair correlation. Additionally, we show in table~\ref{Tab:efficiency} the measured efficiencies for signal and idler retrieved from Eq.~(\ref{eq:efficiencies}) at the different values of the CAR. Due to the long measurement times the coupling efficiencies vary slightly at each CAR.

\begin{figure}[t]
    \centering
    \includegraphics[width=\linewidth]{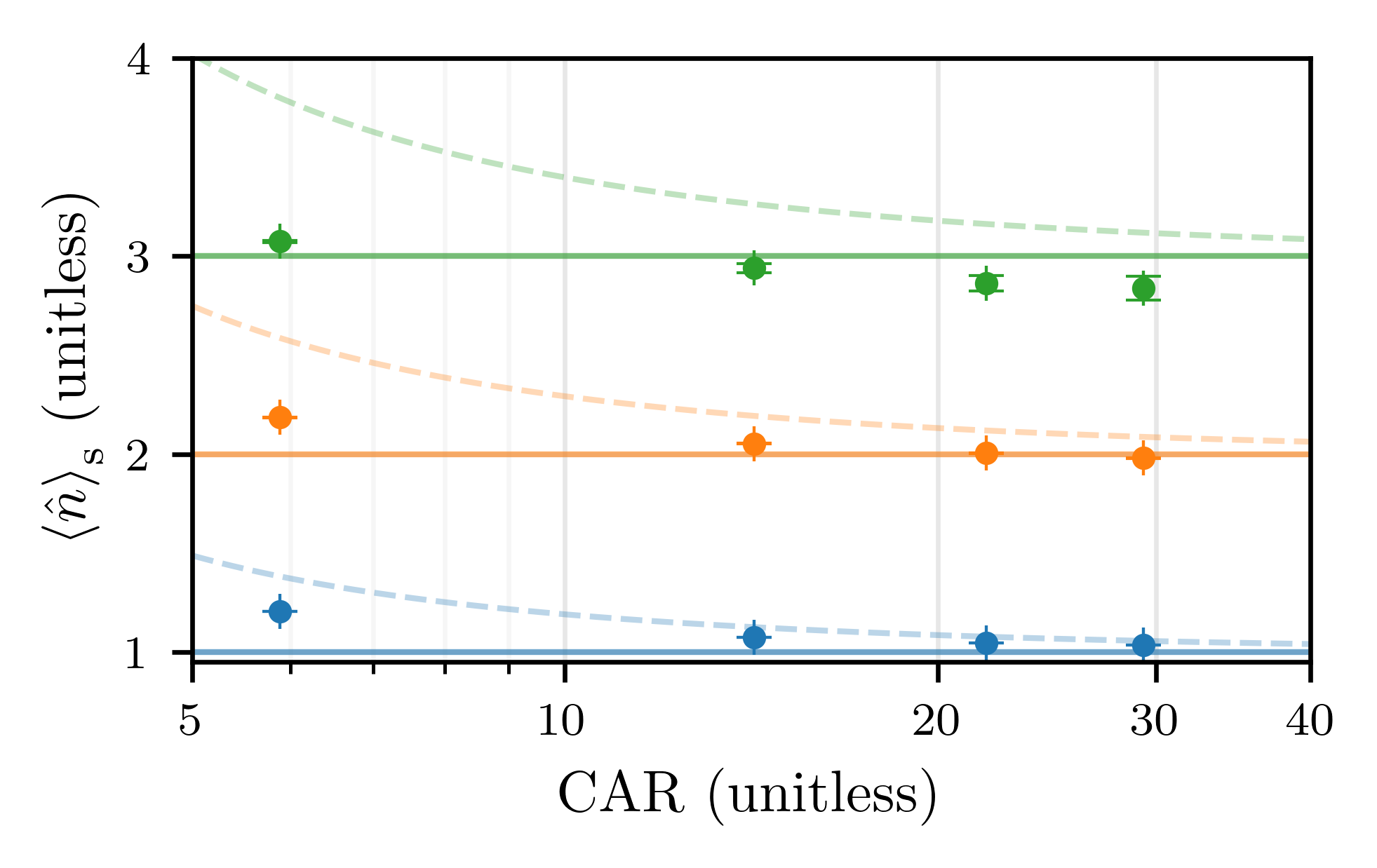}
    \caption{Measured mean photon numbers (symbols) of the heralded states in terms of the CAR. The dashed lines show the theoretical prediction, while the solid lines represent the values for an ideal photon-number states.}
    \label{fig:mean}
\end{figure}

In Fig.~\ref{fig:mean} we depict the experimentally extracted  mean photon numbers of the heralded states retrieved via Eq.~(\ref{eq:mean_exp}) together with the theoretical predictions (see Appendix \ref{sec:app}).  The results for $k=1,2,3$ are depicted in blue, orange and green, respectively. As expected, the evaluated mean photon numbers of the heralded states closely follow the value of $k$. A moderate agreement between the experiment and the theory is achieved at values $\text{CAR}>10$. We believe that the discrepancies at $\text{CAR}<10$ arrive from the fact that our theoretical treatment only holds for single-mode PDC at the low gains \cite{borrero2025advancing}.

\begin{figure}[t]
    \centering
    \includegraphics[width=\linewidth]{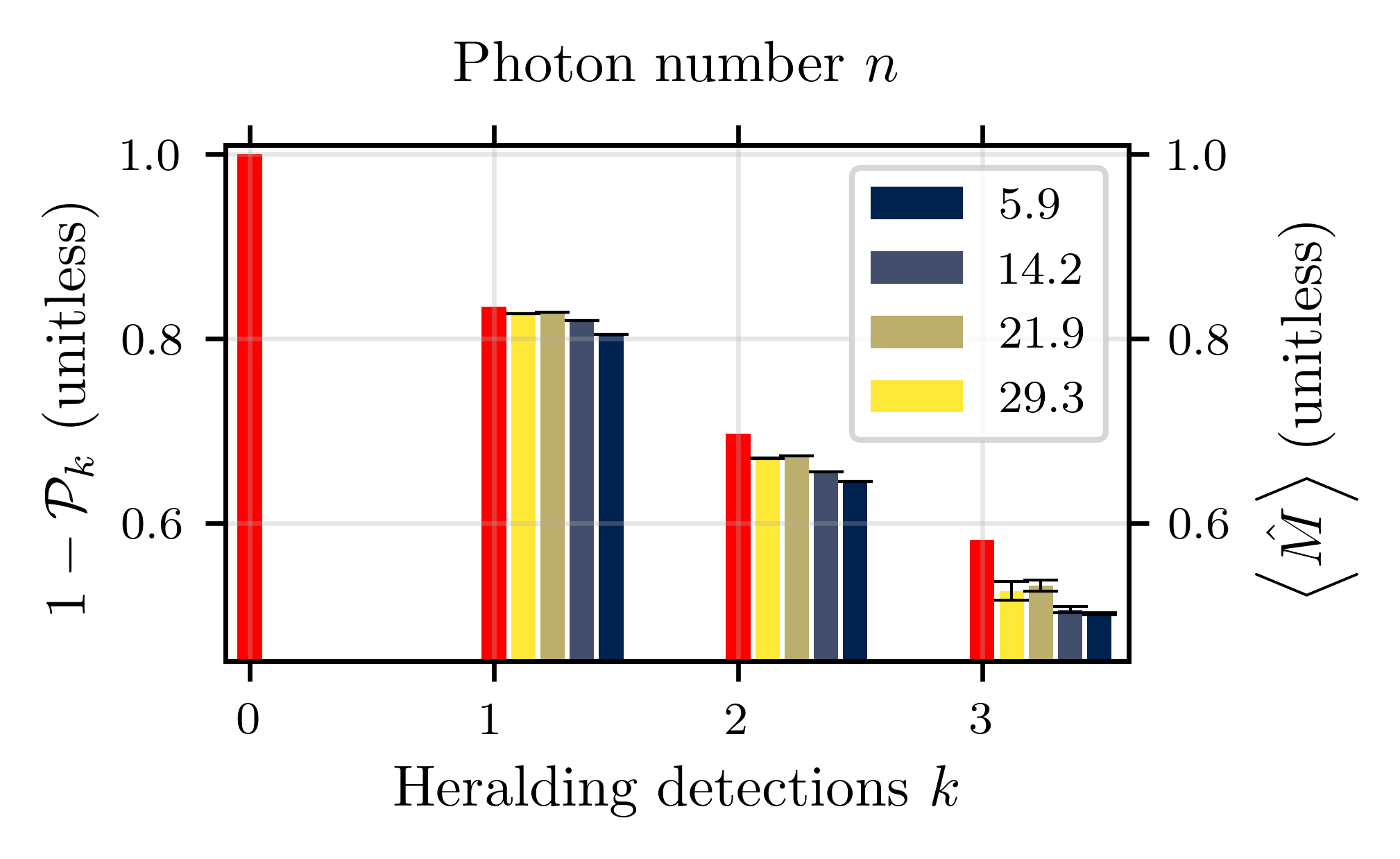}
    \caption{Theoretical (red bars) and experimental expectation values of the moment generating function. Experimental results are evaluated at each measured CAR. The theoretical values for $\big < \hat{M}(\eta)\big>$ are calculated with $\eta=\bar{\mu}_{\text{sc}}=0.165$.}
    \label{fig:prob}
\end{figure}

Next, we probe values of the moment generating function of the heralded $\ket{k}$-states via Eq.~(\ref{eq:Tr_s2}). In Fig.~\ref{fig:prob} we show the experimentally extracted expectation values in terms of $k$ together with a theoretical prediction for an ideal $n$-photon state, for which we expect that $\big < \hat{M}(\eta)\big>=(1-\eta)^n$. At any measured value of the CAR, our results nicely follow the same tendency as expected for the number states, providing us another criterion for validating the features of the heralded states. Besides retrieving values of the moment generating function, the \emph{click} or \emph{no-click} probability  has also been used in the past for retrieving  other specific characteristics of the underlying photon statistics by fitting \cite{Zambra2005,Krapick2014}.

\begin{figure*}[b]
    \centering
    \includegraphics[width= \linewidth]{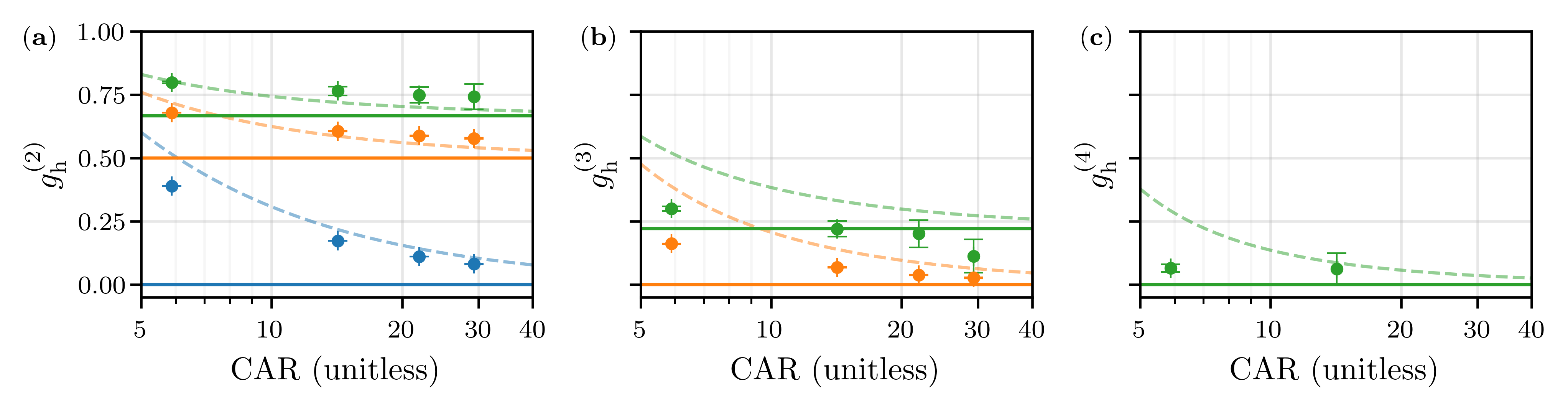}
    \caption{ The measured values $g^{(m)}_{\text{h}}$ $(m = 2,3,4)$ for the heralded  one- (blue), two- (orange) and three-photon (green) states. The dashed line shows the theoretical prediction. The solid lines represent the values expected for ideal photon-number states.}
    \label{fig:g_m_h2}
\end{figure*}

Finally, we study the non-classicality of the heralded states via the normalized factorial moments of the photon number. We measure $g^{(m)}_{\text{h}}(\ket{k})$ for the heralded states up to the order of $m = k+1$ and depict the results in Fig.~\ref{fig:g_m_h2}. Again blue, orange and green are used for $k = 1,2,3$, respectively. The solid lines represent the values expected for ideal photon-number states and the dashed lines are from the theoretical predictions (see Appendix). In Fig.~\ref{fig:g_m_h2}(a) we observe that the measured values of $g^{(2)}_{\text{h}}$  asymptotically approach the ones for ideal $n$-photon states given by $g^{(2)}_{\text{h}}(\ket{n}) = 1-1/n$. Furthermore, we find a good agreement with the theory. 

Further, the results for $g^{(3)}_{\text{h}}$ and $g^{(4)}_{\text{h}}$ are shown in Figs.~\ref{fig:g_m_h2}(b-c), respectively. Again, for the ideal $n$-photon states we expect  $g^{(3)}_{\text{h}}(\ket{n}) = 1-3/n+2/n^2$, while for all investigated states $g^{(4)}_{\text{h}}$ ideally vanish. We observe that the measured values for $g^{(3)}_{\text{h}}$ slightly deviate from the expected ones. Nevertheless, the correct tendencies are observed: The measured values of $g^{(3)}_{\text{h}}$ of the heralded two-photon state tend to approach zero with the increasing CAR and those of the heralded three-photon state are greater than that at all CAR values. Regarding $g^{(4)}_{\text{h}}$, we note that its evaluation involves a simultaneous detection of three photons in idler and four photons in signal, which we obtained only at the two lowest values of CAR.

\begin{figure}
    \centering
    \includegraphics[width=\linewidth]{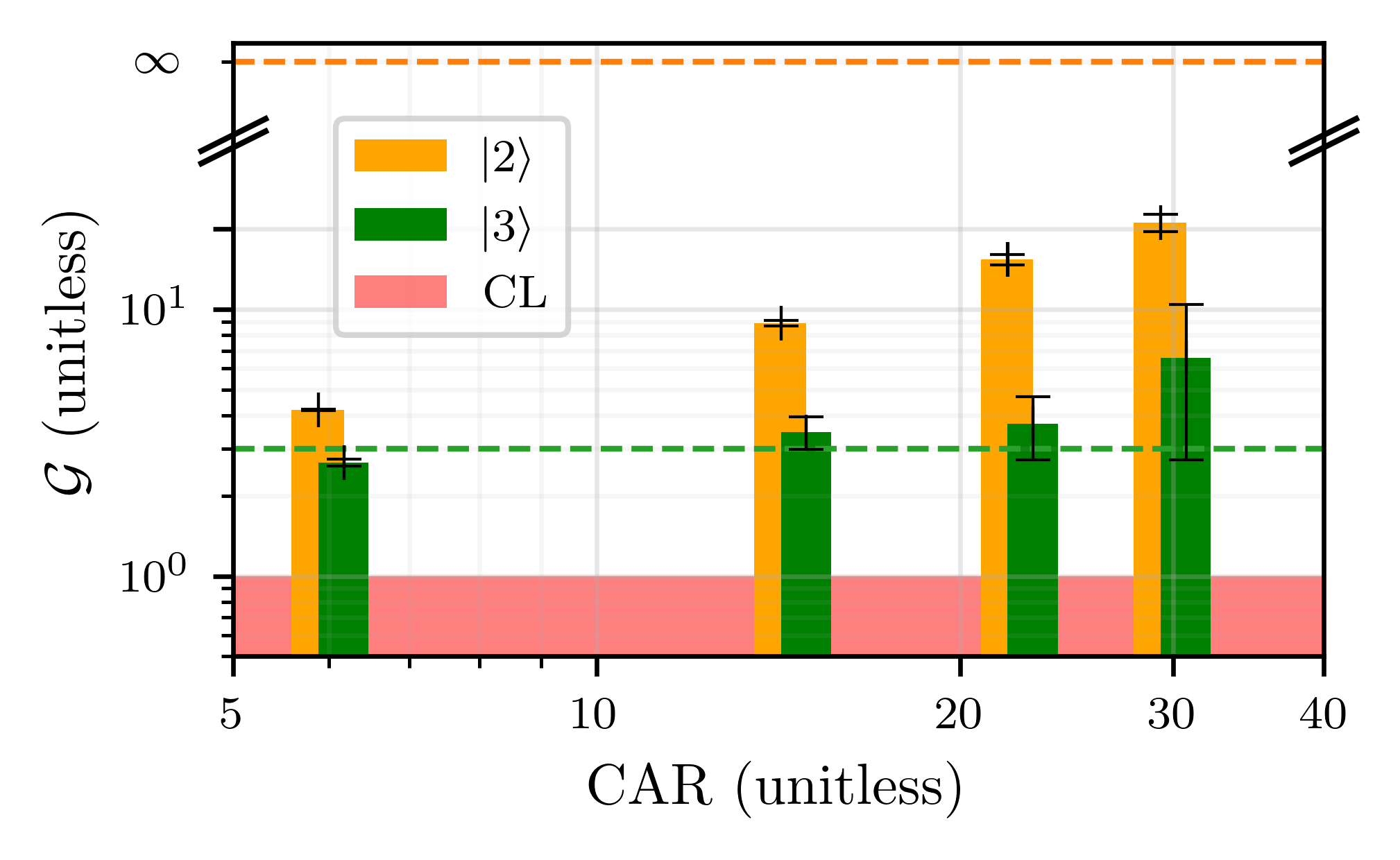}
    \caption{The ratio $\mathcal{G} = g^{(2)}_{\text{h}}$/$g^{(3)}_{\text{h}}$  for the two- and three-photon heralded states in terms of the CAR. The dashed lines represent the ideal values of $\mathcal{G}\rightarrow \infty$ and $\mathcal{G}=3$, respectively. The red-shaded area represents the classical limit.}
    \label{fig:g2g3}
\end{figure}

Most importantly, we observe decreasing values for $g^{(m)}_{\text{h}}$ with increasing $m$, which implies that the condition for classical light $g^{(m+1)}_{\text{h}} \ge g^{(m)}_{\text{h}}$ is violated. As an indication of the higher-order non-classicality we can investigate the ratio $\mathcal{G} = g^{(2)}_{\text{h}}$/$g^{(3)}_{\text{h}}$ for the two- and three-photon heralded states. We depict our results in Fig.~\ref{fig:g2g3}. For an ideal two-photon state one expects that  $\mathcal{G}\rightarrow \infty$. At the highest value of the CAR we maximally reach $\mathcal{G}=$ \SI{21\pm2}{} for the heralded two-photon state. For an ideal three-photon state $\mathcal{G}=3$, while we measured a weighted average of $\mathcal{G}=$\SI{2.9\pm0.2}{}, taking into account weights that are inversely proportional to the error of each individual measurement. Considering that for any classical state $\mathcal{G}\le 1$, our results provide a strong evidence of the higher-order non-classicality.

\onecolumngrid

%\\
\newpage

\twocolumngrid

\emph{Conclusions.~} The implementation of simple and fast methods for the characterization of non-classical light are required for thriving the development of quantum technologies. For this purpose, the measurement of the normalized factorial moments is a suitable tool. We prepare heralded number states up to three-photons in a PP-KTP waveguide in the telecommunication wavelength range. We performed a conventional characterization in terms of the CAR and Klyshko's efficiencies. We measured values of CAR between $5.9$ and $29.3$ and averaged corrected efficiency of $\SI{16.5\pm0.2}{\%}$ and $\SI{23\pm1}{\%}$ for signal and idler, respectively. Regarding the generated heralded states, we retrieved loss-tolerantly their mean photon number just via counting singles and manifold coincidences. Furthermore, we showed this simple tool can also be used for probing values of the moment generating function of the heralded states. Finally, we measured the normalized factorial moments of the heralded state and verified their non-classicality by violating the condition $g^{(m+1)}_{\text{h}} \ge g^{(m)}_{\text{h}} \ge 1$. We observe strong evidence of the higher-order non-classicality of the two- and three-photon heralded states. Our measurement method provides an attractive approach for a precise classification and characterization of multi-photon states.

%TC:ignore
\section{appendix}
\label{sec:app}

For the theoretical predictions we calculate the characteristics of a single-mode twin beam state, which is close to the emission from our PDC source that produces an effective mode-number $<2$ \cite{landazabal2026validating}. The investigated state can be expressed as \cite{migdall2013single}
\begin{equation}
    \ket{\psi} = \sum_{n=0}^{\infty} \lambda^n \sqrt{1-|\lambda|^2} \ket{n,n}_{\text{s,i}}\, 
\label{eq:smTWB}
\end{equation}
in terms of the Fock-states $\ket{n}_{l} (l = s,i)$ in signal (s) and idler (i) and the parameter $\lambda$, which is related to the strength of the squeezing. This model assumes a perfect photon-number correlation between signal and idler. The corresponding density matrix is given by
\begin{equation}
    \hat{\rho}_{\text{s,i}} = \left( 1-|\lambda|^2 \right)\sum_{n,m=0}^{\infty} \lambda^n (\lambda^*)^m  \ket{n,n}_{\text{s,i~i,s}} \bra{m,m} \, .
\label{eq:DM_smTWS}
\end{equation}  
In the single-mode case, the twin beams have a thermal distribution in the photon number basis with mean $\bar{n}$ being related to the $\lambda$-parameter via $|\lambda|^2 = \bar{n}/(1+\bar{n})$ \cite{allevi2022multi}.
We note that we use the signal-idler correlation as a parameter proportional to the created squeezing strength. This can be expressed in terms of the coincidences-to-accidentals ratio (CAR) given for the single-mode twin-beams by CAR $= 2 + 1/\bar{n}$ \cite{borrero2025advancing, laiho2022measuring}.

In the following, we use the idler as herald and extract values for the heralded states' mean photon number and for the normalized factorial moments. For this purpose, we model the time-multiplexed detection, which we experimentally implement, via an equivalent array of multiple detectors (see Fig.~\ref{fig:model}). For the detection system in the heralding arm (idler), we employ the projection \cite{sperling2014quantum}
\begin{eqnarray}
    \hat{O}^{\text{i}}_k &=&  \sum_{m=0}^{k} \binom{N}{k}\binom{k}{m} (-1)^m   \nonumber\\
    & &\times \sum^{\infty}_{n=0}\left(1-\frac{\bar{\mu}_{\text{ic}}}{N}(N+m-k)\right)^{n}\ket{n}_{\text{i~i}}\bra{n} \, ,
\label{eq:POVM_heralding}
\end{eqnarray}
where $N$ is the amount of detection time-bins, $k$ the number of the detected clicks in herald and $\bar{\mu}_{\text{ic}}$ the measured averaged detection efficiency in the idler beam corrected by subtracting the effect of the accidental counts.

By following the treatment in Ref.~\cite{borrero2025advancing} we compute the photon-number content of the heralded beam by taking the trace
$ \Tr_{\text{s}}\left\{ \hat{\rho}_{\text{s}}\ket{n}_{\text{s}~\text{s}}\bra{n}\right\}$, in which the heralded state density matrix in the signal arm takes the form $\hat{\rho}_{\text{s}} = \Tr_{\text{i}}\left\{\hat{O}^{\text{i}}_k \hat{\rho}_{\text{s,i}}\right\} / \Tr_{\text{s,i}}\left\{\hat{O}^{\text{i}}_k \hat{\rho}_{\text{s,i}}\right\}$. This delivers
 \begin{eqnarray}
    p^{\text{s}}_n &=& \mathcal{A} \sum_{m=0}^{k}\binom{N}{k}\binom{k}{m} (-1)^m \frac{\bar{n}^{n}}{ (1+\bar{n})^{n+1}} \nonumber\\
    & &\times\left(1 -\frac{\bar{\mu}_{\text{ic}}}{N}(N+m-k)\right)^n  \, ,
\label{eq:heralded_statistics_2}
\end{eqnarray}
where $\mathcal{A}$ accounts for the normalization factor, so that $\sum_{n}p^{\text{s}}_n = 1$.

The characteristics of the heralded states can be then easily extracted via the photon-number content given in Eq.~(\ref{eq:heralded_statistics_2}). The mean photon number of the heralded state can be then retrieved via
\begin{equation}
     \expval{\hat{n}}_\text{s}^{\text{theory}} = \sum_n n  p^\text{s}_n \, ,
    \label{eq:mean_theory}
\end{equation}
while its normalized $m$-th order normalized factorial moment is gained by \cite{laiho2022measuring}
\begin{equation}
    g^{(m)}_{\text{h,theory}} = \frac{\sum_{n}\frac{n!}{(n-m)!}p^\text{s}_n}{\left(\sum_{n}np^\text{s}_n\right)^m} \, .
\label{eq:gmh_theory}
\end{equation}
For computing the theory curves, we use in Eq.~(\ref{eq:POVM_heralding}) the mean value of $\mu_{\text{ic}}$, which are reported in Table \ref{Tab:efficiency}. Again, this affects the results in Eqs.~(\ref{eq:mean_theory}) and (\ref{eq:gmh_theory}). 

%TC:endignore

% main text: 2046 words 

% captions: 234 words

% inline equations: 119 words

% equations: 7 Eqs.*16 words/Eq. = 112 Words

% table: 13 + 6.5words/line * 3 lines = 33 Words

% Figure 1: 150/0.91 + 20 : 185
% Figure 2: 150/2.2 + 20 : 88
% Figure 3: 150/0.81 + 20 : 113
% Figure 4: 150/1.6 + 20 : 113
% Figure 5: 150/1.6 + 20 : 113
% Figure 6: 600/3.83 + 40: 197
% Figure 7: 150/1.6 + 20: 113
%Total figures: 922 words

% Actual Total: 3466 words

% Limit: 3750 words

\bibliography{references.bib}

\end{document}